\documentclass{article}

\usepackage{arxiv}

\usepackage[utf8]{inputenc} 
\usepackage[T1]{fontenc}    

\usepackage{amsmath,amssymb,amsthm,mathtools,bm}
\usepackage{nicefrac}       
\usepackage{scalerel,stackengine}
\usepackage{graphicx}
\usepackage{booktabs}
\usepackage{siunitx}
\usepackage{doi} 
\usepackage{enumitem}
\usepackage{hyperref}
\usepackage{url}            
\usepackage[nameinlink,noabbrev]{cleveref}
\usepackage{microtype}
\usepackage{siunitx}

\usepackage{matlab-prettifier}
\usepackage{listings}

\usepackage{titlesec}
\titleformat{\paragraph}
{\normalfont\normalsize\bfseries}{\theparagraph}{1em}{}
\titlespacing*{\paragraph}
{0pt}{3.25ex plus 1ex minus .2ex}{1.5ex plus .2ex}

\usepackage{cleveref}
\crefname{figure}{Figure}{Figures}
\crefname{table}{Table}{Tables}
\crefname{section}{}{}
\crefname{subsection}{Section}{Sections}
\crefname{subsubsection}{Section}{Sections}
\crefname{equation}{Equation}{Equations}
\crefname{algorithm}{Algorithm}{Algorithms}
\usepackage{etoolbox}
\crefname{section}{}{}

\usepackage[font=scriptsize]{caption}
\hypersetup{
pdftitle={Data-adaptive spline surfaces for non-separable hyperelastic energy functions},
pdfsubject={cond-mat.mtrl-sci},
pdfauthor={Simon Wiesheier, Miguel Angel Moreno-Mateos, Paul Steinmann},
pdfkeywords={Data-driven methods, B-splines, Automated model discovery, Interpolation, Hyperelasticity},
}

\hypersetup{
    colorlinks=true,
    linkcolor=blue,     
    filecolor=magenta,  
    urlcolor=black,      
    citecolor=blue,    
}

\newcommand\equalhat{\mathrel{\stackon[1.5pt]{=}{\stretchto{%
    \scalerel*[\widthof{=}]{\wedge}{\rule{1ex}{3ex}}}{0.5ex}}}}
\newcommand{\R}{\mathbb{R}}
\newcommand{\tr}{\operatorname{tr}}
\newcommand{\cof}{\operatorname{cof}}
\newcommand{\diag}{\operatorname{diag}}

\newcommand{\F}{\mathbf{F}}
\newcommand{\C}{\mathbf{C}}

\newcommand{\Pnom}{\mathbf{P}}

\newcommand{\Ione}{{\bar{I}_1}}
\newcommand{\Itwo}{{\bar{I}_2}}
\newcommand{\Ithree}{{\bar{I}_3}}

\newcommand{\IoneUT}{\bar{I}_{1\ut}}
\newcommand{\IoneBT}{\bar{I}_{1\bt}}
\newcommand{\IonePS}{\bar{I}_{1\ps}}
\newcommand{\ItwoUT}{\bar{I}_{2\ut}}
\newcommand{\ItwoBT}{\bar{I}_{2\bt}}
\newcommand{\ItwoPS}{\bar{I}_{2\ps}}

\newcommand{\W}{W}
\newcommand{\Wsep}{W_{\mathrm{sep}}}
\newcommand{\Wsurf}{W_{\mathrm{surf}}}

\newcommand{\ut}{\mathrm{UT}}
\newcommand{\bt}{\mathrm{BT}}
\newcommand{\ps}{\mathrm{PS}}

\newcommand{\xiu}{\xi}
\newcommand{\xiv}{\eta}

\newcommand{\Nbu}[1]{N_{#1}}
\newcommand{\Nbv}[1]{M_{#1}}

\newcommand{\Ivec}{\boldsymbol \iota}

\newcommand{\calD}{\mathcal{D}}
\newcommand{\calA}{\mathcal{A}}
\newcommand{\calL}{\mathcal{L}}
\newcommand{\calR}{\mathcal{R}}

\newcommand{\p}{\bm{\theta}}

\theoremstyle{plain}
\newtheorem{remark}{Remark}

\begin{document}

\title{Data-adaptive spline surfaces for non-separable hyperelastic energy functions}

\author{ 
    \href{https://orcid.org/0009-0008-9625-3446}{\includegraphics[scale=0.06]{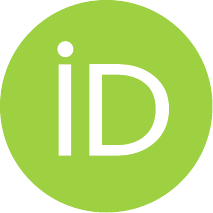}\hspace{1mm}Simon Wiesheier}\thanks{Corresponding author: simon.wiesheier@fau.de} \\
	Institute of Applied Mechanics\\
	Friedrich-Alexander-Universität Erlangen–Nürnberg\\
	91058, Erlangen, Germany\\
    \And
    \href{https://orcid.org/0000-0002-3476-2180}{\includegraphics[scale=0.06]{orcid.pdf}\hspace{1mm}Miguel Angel Moreno-Mateos}\thanks{Corresponding author: miguel.moreno@fau.de} \\
	Institute of Applied Mechanics\\
	Friedrich-Alexander-Universität Erlangen–Nürnberg\\
	91058, Erlangen, Germany\\
	\texttt{miguel.moreno@fau.de} \\
	\And
	\href{https://orcid.org/0000-0003-1490-947X}{\includegraphics[scale=0.06]{orcid.pdf}\hspace{1mm}Paul Steinmann} \\
	Institute of Applied Mechanics\\
	Friedrich-Alexander-Universität Erlangen–Nürnberg\\
	91058, Erlangen, Germany\\
	Glasgow Computational Engineering Centre\\
	University of Glasgow\\
	G12 8QQ, United Kingdom\\
}

\maketitle

\begin{abstract}
Invariant-based models for incompressible isotropic hyperelasticity are typically formulated as functions of the first and second invariants, $W = W(\Ione, \Itwo)$. A widely used class of models employs separable representations of the form $W(\Ione, \Itwo) = W_1(\Ione) + W_2(\Itwo)$, which enable efficient calibration and straightforward enforcement of modeling constraints. However, this decomposition implicitly restricts the coupling between the invariants and may limit the achievable accuracy for complex material responses.
Fully coupled data-driven approaches overcome this limitation but often require nonlinear optimization and large parameter sets. In this contribution, we propose a compact alternative: a bivariate B-spline surface defined directly on the physically admissible invariant domain. By aligning the approximation space with physically realizable states, all model parameters contribute meaningfully to the constitutive response.
We utilize homogeneous deformation modes to perform a calibration directly from analytical stress relations, eliminating the need for finite element model updating. Owing to the linear dependence of the spline representation on its coefficients, the resulting parameter identification problem reduces to a constrained linear least-squares problem. This enables fast, robust, and initialization-independent calibration, which makes parameter identification practically instantaneous.
The results demonstrate that the proposed model improves accuracy compared to separable approaches while requiring only mild regularization in weakly sampled regions. The combination of computational efficiency and the linear structure of a highly expressive spline surface makes the approach particularly attractive for applications requiring repeated calibration, such as uncertainty quantification or interactive material characterization.
\end{abstract}

\keywords{Data-driven methods $|$ B-splines $|$ Automated model discovery $|$ Interpolation $|$ Hyperelasticity}

\section{Introduction}

Understanding and modeling the mechanical behavior of materials requires constitutive descriptions that accurately capture the response under a variety of loading conditions. In the context of hyperelasticity, a material class describing finite elastic deformations, the constitutive response is fully characterized by a scalar strain-energy density function \cite{holzapfel_nonlinear_2001}. 
For isotropic and incompressible materials, this function can be expressed in terms of two invariants as
\begin{align}
  \W = \W(\Ione,\Itwo),
\end{align}
where $\Ione$ and $\Itwo$ denote the first and second invariants of the isochoric
right Cauchy--Green tensor.

Classical invariant-based models (see, e.g., \cite{steinmann_hyperelastic_2012,ricker_systematic_2023}),
such as neo-Hookean, Mooney--Rivlin \cite{mooney_theory_1940}, or Yeoh \cite{yeoh_characterization_1990}, prescribe a fixed functional form for $\W$ in advance. While this is attractive from a modeling standpoint, it can be restrictive when the objective is to infer constitutive behavior directly from experimental data.

To address this limitation, a range of data-driven approaches have been proposed, including neural-network models \cite{fuhg_review_2025,kalina_automated_2022,klein_polyconvex_2022,klein_neural_2025,linden_neural_2023},
Gaussian processes \cite{frankel_tensor_2019}, symbolic regression \cite{abdusalamov_automatic_2023}, and neural ordinary differential equations \cite{tac_data-driven_2023}. Spline functions have also recently been employed as activation functions in neural-network architectures, for example, in Kolmogorov--Arnold networks (KANs) \cite{thakolkaran_can_2025,abdolazizi_constitutive_2025}. Another emerging direction is material fingerprinting \cite{flaschel_unsupervised_2026}, where precomputed constitutive response databases enable fast model identification without online optimization.

Among these developments, spline-based approaches (\cite{sussman_model_2009, dal_data-driven_2023,tikenogullari_data-driven_2023} offer an attractive compromise between flexibility and interpretability. In previous works by the authors \cite{wiesheier_discrete_2023,wiesheier_versatile_2024,moreno-mateos_biaxial_2025,wiesheier_data-adaptive_2026},
a data-adaptive B-spline framework \cite{piegl_nurbs_1996,d_boor_practical_1978} was developed to represent constitutive functions by discretizing the invariant space and interpolating the strain-energy density. The interpolation values serve as material parameters and are identified from experimental data, typically using global reaction forces and displacement fields obtained from digital image correlation.

Initial applications of this framework focused on hyperelastic responses of soft materials. Subsequent extensions addressed data-adaptive configurational forces \cite{moreno-mateos_biaxial_2025} and viscoelastic behavior through a data-adaptive Generalized Standard Material formulation (cf. the DAVIS framework in \cite{wiesheier_data-adaptive_2026}). Moreover, modeling constraints such as polyconvexity \cite{hartmann_polyconvexity_2003,schroder_invariant_2003} or a stress-free reference configuration can be incorporated directly into the spline construction.

Parameter identification typically requires solving an optimization problem, which may become computationally demanding for complex models. In contrast, spline-based representations offer a key structural advantage: B-spline interpolation is linear in its coefficients. When combined with homogeneous deformation protocols with analytical stress relations, calibration reduces to a constrained linear least-squares problem. As a result, the optimal parameters can be determined efficiently without requiring an initial guess, and the solution is unique under mild conditions. At the same time, the representation remains flexible and data-adaptive.

In the aforementioned works, the strain-energy density was represented in separable form as
\begin{align}
  \Wsep(\Ione,\Itwo)=W_1(\Ione)+W_2(\Itwo),
\end{align}
where $W_1$ and $W_2$ are univariate spline functions. Although this decomposition
enables efficient calibration and straightforward enforcement of constraints, it
implicitly assumes separability of the invariant contributions.

The present contribution lifts this restriction while preserving the favorable linear structure of the calibration problem. This is achieved by (i) restricting attention to homogeneous deformation modes, enabling calibration based on analytical stress relations, (ii) exploiting the fact that only a subset of the $(\Ione,\Itwo)$-plane is physically admissible for incompressible deformations, and (iii) introducing a tensor-product spline surface defined on a mapped version of this admissible domain. 

The admissible invariant domain is bounded by the uniaxial and equi-biaxial branches, so that a direct tensor-product spline representation in $(\Ione,\Itwo)$ would allocate interpolation support in non-physical regions. To avoid this, we introduce a coordinate transformation that maps the admissible domain onto the unit square. This yields a compact and efficient spline representation with full coupling between the invariants, analytical derivatives, and a reduced number of parameters.

\begin{figure}
    \centering
    \includegraphics[width=0.99\linewidth]{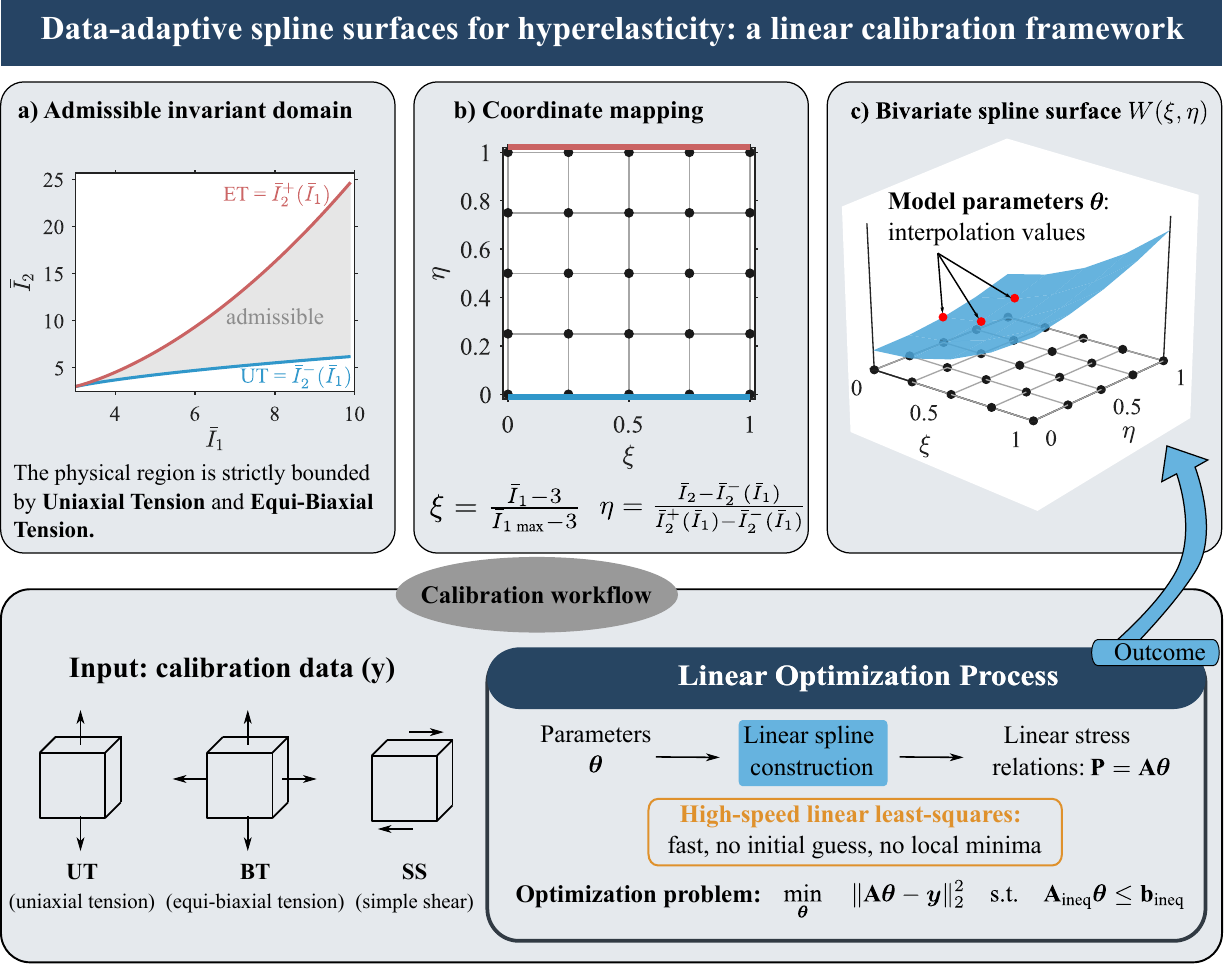}
    \caption{The admissible invariant domain (a) in the ($\Ione,\Itwo$) space is mapped into a normalized ($\xiu,\xiv$) space using the boundary curves. This space is discretized (b), and a tensor-product spline surface (c) is defined. The interpolation values are calibrated using experimental data from homogeneous deformation modes. Owing to the linearity of spline construction, calibration reduces to a linear least-squares problem, enabling rapid and robust model discoevery.}
    \label{fig:workflow}
\end{figure}

The resulting framework combines the expressive power of fully coupled constitutive models with the computational efficiency of linear least-squares calibration. The proposed workflow is illustrated in ~\cref{fig:workflow}.

This paper proceeds as follows: in section~\cref{sec:kinematics}, we introduce the relevant kinematics and analytical stress--stretch relations. In section~\cref{sec:data_adaptive_adm_domain}, we present the proposed spline surface on the admissible invariant domain and the associated coordinate mapping. This mapping is essential because a direct tensor-product spline representation in $(\Ione,\Itwo)$ space would allocate interpolation support to non-physical regions. We then formulate the resulting constrained linear least-squares calibration problem. Finally, in section~\cref{sec:results}, we benchmark the proposed approach against separable models and direct representations in the full $(\Ione,\Itwo)$ space and report computational efficiency.

\section{Kinematics and homogeneous deformation modes}
\label{sec:kinematics}

Let $\F \in \R^{3\times 3}$ denote the deformation gradient, $J=\det \F > 0$
its determinant, and
\begin{align}
  \C = \F^\top \F,
  \qquad
  \bar{\C}=J^{-2/3}\C .
\end{align}

Following \cite{flory_thermodynamic_1961}, $\bar{\C}$ denotes the isochoric part of the right Cauchy--Green tensor $\C$ obtained by the volumetric--isochoric decomposition of the deformation gradient. For incompressible deformation,
$J=1$ so that $\bar{\C}=\C$.

We employ the classical isochoric invariants
\begin{align}
  \Ione = \tr(\bar{\C}) = \kappa_1 + \kappa_2 + \kappa_3 ,
  \qquad
  \Itwo = \tr(\cof \bar{\C}) = \kappa_1\kappa_2 + \kappa_2\kappa_3 + \kappa_1\kappa_3 ,
\end{align}
which, together with $\Ithree=\det(\bar{\C})=\kappa_1\kappa_2\kappa_3=1$, appear as coefficients of the characteristic polynomial
\begin{align}
  \kappa^3-\Ione\kappa^2+\Itwo\kappa-1=0,
  \label{eq:CE}
\end{align}
with $\kappa$ the eigenvalues of $\bar{\C}$. 

The strain-energy density is written as
\begin{align}
  \Psi(\F,p)=\W(\Ione,\Itwo)-p[J-1],
\end{align}
where $p$ denotes the hydrostatic pressure acting as a Lagrange multiplier for the incompressibility constraint. The Piola stress then follows as
\begin{align}
  \Pnom=\frac{\partial\W}{\partial\F}-p\F^{-\top}.
\end{align}

For homogeneous deformations, the stress can be expressed in terms of the invariant derivatives
\begin{align}
  \W_{,1}=\frac{\partial\W}{\partial\Ione},
  \qquad
  \W_{,2}=\frac{\partial\W}{\partial\Itwo}.
\end{align}

The hydrostatic pressure $p$ is determined from the plane-stress condition in
thickness direction,
\begin{align}
  P_{33}=0,
\end{align}
which uniquely fixes $p$ for the deformation states considered below. The resulting nominal stress tensor $\Pnom$ is then fully determined, and the relevant stress component can be written in closed form for each loading mode.

\paragraph{Homogeneous deformation modes}

We consider three classical deformation modes: uniaxial tension ($\ut$), equi-biaxial tension ($\bt$), and pure shear ($\ps$). For each mode, the deformation gradient defines analytical expressions for the
invariants $(\Ione,\Itwo)$ and for the measured nominal stress component.

\paragraph{Uniaxial tension ($\ut$)}
For stretch $\lambda>0$
\begin{align}
  \F_{\ut} \equalhat \diag(\lambda,\lambda^{-1/2},\lambda^{-1/2}),
\end{align}
which yields
\begin{align}
  \IoneUT=\lambda^2+2\lambda^{-1},
  \qquad
  \ItwoUT=\lambda^{-2}+2\lambda .
  \label{eq:Iut}
\end{align}

The nominal stress tensor is diagonal, and after elimination of the pressure, the only nonzero component is the loading component $P_{11}$. It reads
\begin{align}
  P_{\ut}(\lambda)
  =
  2\Big[[\lambda-\lambda^{-2}]\W_{,1}
  +[1-\lambda^{-3}]\W_{,2}\Big].
  \label{eq:Put}
\end{align}

\paragraph{Equi-biaxial tension ($\bt$)}
For
\begin{align}
  \F_{\bt} \equalhat \diag(\lambda,\lambda,\lambda^{-2}),
\end{align}
one obtains
\begin{align}
  \IoneBT=2\lambda^2+\lambda^{-4},
  \qquad
  \ItwoBT=2\lambda^{-2}+\lambda^4.
  \label{eq:Ibt}
\end{align}

The nominal stress tensor is again diagonal. By symmetry, the two in-plane components coincide, $P_{11}=P_{22}$, while $P_{33}=0$ by construction. The measured nominal stress is therefore
\begin{align}
  P_{\bt}(\lambda)
  =
  2\Big[[\lambda-\lambda^{-5}]\W_{,1}
  +[\lambda^3-\lambda^{-3}]\W_{,2}\Big].
  \label{eq:Pbt}
\end{align}

\paragraph{Pure shear ($\ps$)}
For
\begin{align}
  \F_{\ps} \equalhat \diag(\lambda,1,\lambda^{-1}),
\end{align}
the invariants become
\begin{align}
  \IonePS=\lambda^2+1+\lambda^{-2},
  \qquad
  \ItwoPS=\lambda^{-2}+1+\lambda^2.
  \label{eq:Ips}
\end{align}
The nominal stress tensor is diagonal, and after pressure elimination $P_{11}$ and $P_{22}$ remain nonzero. In particular, $P_{11}$ reads
\begin{align}
  P_{\ps}(\lambda)
  =
  2\left[\lambda-\lambda^{-3}\right] \Big[\W_{,1} + \W_{,2}\Big].
\end{align}

\section{Data-adaptive spline surface on the admissible domain}
\label{sec:data_adaptive_adm_domain}

The key objective of the present work is to construct a compact bivariate spline representation for the strain-energy density that allows for coupled dependence on $\Ione$ and $\Itwo$ while preserving the favorable linear structure of the calibration problem. However, a direct tensor-product spline surface in the full $(\Ione,\Itwo)$ space is not well aligned with the kinematics of incompressible isotropic deformations: only a proper subset of that plane is physically attainable. As a consequence, a direct discretization would place interpolation points and spline support also in inadmissible regions, where no deformation state exists, and no calibration data can occur. This leads to inefficient use of parameters and weakly constrained spline coefficients. We therefore first characterize the admissible invariant domain and introduce a coordinate mapping that transforms it to a unit square parameter space on which the tensor-product spline surface can be defined efficiently. The mapped representation then serves as the basis for the constrained linear least-squares calibration.

\subsection{Admissible invariant domain and coordinate mapping}
\label{sec:mapping}

We begin by characterizing the subset of the invariant space that is physically attainable under incompressible deformations. This admissible domain provides the geometric foundation for the construction of the spline and clarifies why a direct discretization of the tensor-product in the full $(\Ione,\Itwo)$ plane is inefficient. The following construction is in line with earlier characterizations of the admissible invariant set for isochoric deformations \cite{sawyers_possible_1977,dammas_when_2025}.

\paragraph{Definition}

Not every pair $(\Ione,\Itwo)\in\R^2$ corresponds to a physically realizable incompressible deformation. The admissible set is a proper subset
\begin{align}
  \calA
  =
  \big\{
  (\Ione,\Itwo)\in [3,\infty)\times[3,\infty)
  \,\big|\,
  \Itwo^{-}(\Ione)\le \Itwo \le \Itwo^{+}(\Ione)
  \big\},
\end{align}
where $\Itwo^{-}(\Ione)$ and $\Itwo^{+}(\Ione)$ denote the lower and upper bounds of $\Itwo$ for a given $\Ione$. These bounds are attained by uniaxial and equi-biaxial deformations, respectively.

The corresponding boundary curves can be expressed in parametric form as
\begin{align}
  \Gamma^{-}
  &=
  \Big\{
  \big[\lambda^2+2\lambda^{-1},\lambda^{-2}+2\lambda\big]
  \; ; \; \lambda>0
  \Big\},
  \\
  \Gamma^{+}
  &=
  \Big\{
  \big[2\lambda^2+\lambda^{-4},2\lambda^{-2}+\lambda^4\big]
  \; ; \; \lambda>0
  \Big\},
\end{align}
which represent the graphs of $\Itwo^{-}(\Ione)$ and $\Itwo^{+}(\Ione)$ in parametric form.

\begin{remark}
A crude necessary estimate of the admissible set is
\begin{align}
  \sqrt{3\Ione}\le \Itwo \le \frac{\Ione^2}{3},
\end{align}
but the sharp boundary is given by the uniaxial and equi-biaxial branches above.
\end{remark}

\paragraph{Coordinate mapping of the admissible domain}
\label{sec:mapping}

To construct a tensor-product spline surface without wasting interpolation support outside $\calA$, we introduce coordinates that straighten the admissible domain into a unit square. In this way, all spline degrees of freedom are assigned to physically realizable states, and the resulting parameterization remains compact. We define
\begin{align}
  u &= \Ione,
  \\
  v &= \frac{\Itwo-\Itwo^{-}(\Ione)}{\Itwo^{+}(\Ione)-\Itwo^{-}(\Ione)}.
  \label{eq:vmap}
\end{align}

Subsequently,
\begin{align}
  \Itwo = \Itwo^{-}(u) + v\big[\Itwo^{+}(u)-\Itwo^{-}(u)\big],
\end{align}
and the admissible domain is mapped to the rectangle
\begin{align}
  (u,v)\in [u_{\min},u_{\max}]\times[0,1].
\end{align}

Here $u_{\min}=3$, which is the lower bound of the first invariant $\Ione$, while $u_{\max}$ is chosen from the largest invariant value covered by the calibration data and can be computed using ~\cref{eq:Iut,eq:Ibt,eq:Ips}.

For numerical convenience, we additionally normalize $u$ affinely to
\begin{align}
  \xiu = \frac{u-u_{\min}}{u_{\max}-u_{\min}},
  \qquad
  \xiv = v,
\end{align}
so that $(\xiu,\xiv)\in[0,1]^2$.

The required chain rule for the nominal stress calculation is explicit. Let the strain-energy density function be
\begin{align}
  \W(\Ione,\Itwo)=\widehat{\W}(\xiu(\Ione),\xiv(\Ione,\Itwo)).
\end{align}
Then the partial derivatives with respect to the invariants are given by
\begin{align}
  \W_{,2}
  =
  \widehat{\W}_{,\xiv}
  \frac{\partial \xiv}{\partial \Itwo},
  \qquad
  \W_{,1}
  =
  \widehat{\W}_{,\xiu}\frac{1}{u_{\max}-u_{\min}}
  +
  \widehat{\W}_{,\xiv}\frac{\partial \xiv}{\partial \Ione}.
  \label{eq:chainrule}
\end{align}

To evaluate these quantities, we compute the derivatives of the coordinate map.  
Introducing the admissible width
\begin{align}
  \Delta(\Ione):=\Itwo^{+}(\Ione)-\Itwo^{-}(\Ione),
\end{align}
the definition \eqref{eq:vmap} yields
\begin{align}
  \frac{\partial \xiv}{\partial \Itwo}
  =
  \frac{1}{\Delta(\Ione)}.
  \label{eq:dv_dI2}
\end{align}
Differentiating \eqref{eq:vmap} with respect to $\Ione$ gives
\begin{align}
  \frac{\partial \xiv}{\partial \Ione}
  =
  \frac{-\Itwo^{-\,\prime}(\Ione)\,\Delta(\Ione)
        -[\Itwo-\Itwo^{-}(\Ione)]\,\Delta'(\Ione)}
       {\Delta(\Ione)^2},
  \label{eq:dv_dI1}
\end{align}
where
\begin{align}
  \Delta'(\Ione)=\Itwo^{+\,\prime}(\Ione)-\Itwo^{-\,\prime}(\Ione).
\end{align}

Hence the invariant derivatives required in the stress relations can be written compactly as
\begin{align}
  \W_{,1}
  &=
  \frac{\widehat{\W}_{,\xiu}}{u_{\max}-u_{\min}}
  +
  \widehat{\W}_{,\xiv}\frac{\partial \xiv}{\partial \Ione},
  \label{eq:dWd1}
  \\
  \W_{,2}
  &=
  \widehat{\W}_{,\xiv}\frac{1}{\Delta(\Ione)}.
  \label{eq:dWd2}
\end{align}

The remaining ingredients are the boundary functions $\Itwo^{-}(\Ione)$ and $\Itwo^{+}(\Ione)$ and their derivatives. These functions describe the sharp boundary of the admissible invariant domain. Instead of evaluating their closed-form expressions, which are presented in equations 30 and 31 in \cite{dammas_when_2025}, we compute them numerically from the discriminant relation of the characteristic \cref{eq:CE}. Boundary points satisfy the cubic equation
\begin{align}
  C(\Ione,\Itwo)
  :=
  \Itwo^3
  - \frac{1}{4}\Ione^2\Itwo^2
  - \frac{9}{2}\Ione\Itwo
  + \Ione^3
  + \frac{27}{4}
  =0,
  \label{eq:cubicI2}
\end{align}
which corresponds to the discriminant condition of the admissible invariant set.  
For fixed $\Ione\ge3$, the smallest and largest admissible real roots of \eqref{eq:cubicI2} define
\begin{align}
  \Itwo^{-}(\Ione),
  \qquad
  \Itwo^{+}(\Ione).
\end{align}

Their derivatives follow by implicit differentiation of \eqref{eq:cubicI2}.  
Since $P(\Ione,\Itwo^\pm(\Ione))=0$, differentiation yields
\begin{align}
  \Itwo^{\pm\,\prime}(\Ione)
  =
  -\frac{\partial_{\Ione}P(\Ione,\Itwo^\pm)}
         {\partial_{\Itwo}P(\Ione,\Itwo^\pm)},
\end{align}
with
\begin{align}
  \partial_{\Itwo}P(\Ione,\Itwo)
  &= 3\Itwo^2 - \frac{1}{2}\Ione^2\Itwo - \frac{9}{2}\Ione,
  \\
  \partial_{\Ione}P(\Ione,\Itwo)
  &= -\frac{1}{2}\Ione\Itwo^2 - \frac{9}{2}\Itwo + 3\Ione^2 .
\end{align}

\begin{remark}[Degeneracy at the undeformed configuration]
At the undeformed state $(\Ione,\Itwo)=(3,3)$ the admissible invariant domain collapses to a single point, implying
\begin{align}
  \Delta(3)=\Itwo^{+}(3)-\Itwo^{-}(3)=0.
\end{align}
Consequently, the normalization in \eqref{eq:vmap} becomes singular at $\Ione=3$ and numerically ill-conditioned in its vicinity. In the numerical implementation, we therefore replace $\Delta(\Ione)$ by a slightly regularized width
\begin{align}
  \Delta_{\mathrm{eff}}(\Ione)
  =
  \sqrt{\Delta(\Ione)^2+\delta^2},
\end{align}
with a small parameter $\delta>0$. This opens the admissible band near the apex of the invariant domain while leaving the mapping essentially unchanged away from the undeformed configuration.
\end{remark}

\begin{remark}[Polyconvex variant of the mapping]
In our numerical implementation, the mapping is performed using the polyconvex second invariant \cite{ball_convexity_1976, hartmann_polyconvexity_2003}
\begin{align}
  \widetilde{\Itwo} = \Itwo^{3/2}-3\sqrt{3}.
  \label{eq:I2poly}
\end{align}
In that case, the normalized coordinate is not formed from $\Itwo$ directly,
but from the transformed quantity and the correspondingly transformed bounds,
i.e.,
\begin{align}
  v
  =
  \frac{\widetilde{\Itwo}-\widetilde \Itwo^{-}(\Ione)}
       {\Delta_{\mathrm{eff}}(\Ione)},
  \qquad
  \widetilde \Itwo^{\pm}(\Ione)
  =
  \big(\Itwo^{\pm}(\Ione)\big)^{3/2}-3\sqrt{3},
\end{align}
with regularized width
\begin{align}
  \Delta_{\mathrm{eff}}(\Ione)
  =
  \sqrt{
    \left[\widetilde \Itwo^{+}(\Ione)-\widetilde \Itwo^{-}(\Ione)\right]^2
    +\delta^2 }.
\end{align}
\end{remark}

\subsection{Bivariate spline surface}
\label{sec:spline_surface}

Having mapped the admissible invariant domain to a unit square parameter space, we can now define the spline representation in a form that is both compact and directly compatible with standard tensor-product interpolation.

Earlier works by the authors employed separable spline representations of the form
\begin{align}
  \W(\Ione,\Itwo)=\W_1(\Ione)+\W_2(\Itwo),
\end{align}
where $W_1$ and $W_2$ denote univariate B-spline interpolants calibrated from experimental data
\cite{wiesheier_discrete_2023,wiesheier_versatile_2024,wiesheier_data-adaptive_2026}. While this representation is transparent and computationally efficient, it implicitly assumes that the influence of the invariants is additive.

To remove the separability restriction, we consider a tensor-product spline surface defined on the mapped coordinates:
\begin{align}
  \Wsurf(\Ione,\Itwo)
  =
  \widehat{\W}(\xiu,\xiv)
  =
  \sum_{i=1}^{n_\xiu}\sum_{j=1}^{n_\xiv}
  c_{ij}\,\Nbu{i}(\xiu)\,\Nbv{j}(\xiv),
  \label{eq:surface}
\end{align}
where $\Nbu{i}$ and $\Nbv{j}$ denote cubic B-spline basis functions on $[0,1]$.

Rather than optimizing the B-spline control points $c_{ij}$ directly, we prescribe interpolation sites $(\xiu_i,\xiv_j)\subset[0,1]^2$ and treat the interpolation values
\begin{align}
  \iota_{ij}=\W(\xiu_i,\xiv_j)
\end{align}
as the model parameters.

Together with the interpolation sites, these values uniquely determine the spline surface. In particular, the control points $c_{ij}$ are obtained by solving a linear interpolation system that enforces
\begin{align}
  \sum_{i=1}^{n_\xiu}\sum_{j=1}^{n_\xiv}
  c_{ij}\,\Nbu{i}(\xiu_k)\,\Nbv{j}(\xiv_l)
  =
  \iota_{kl}
  \qquad \forall\ \  k,l.
\end{align}

Collecting the interpolation values in the matrix $\Ivec\in\R^{n_\xiu\times n_\xiv}$, the parameter vector is defined as
\begin{align}
  \p=\operatorname{vec}(\Ivec).
\end{align}

Since the spline construction is linear in the interpolation values, the resulting surface and its derivatives depend linearly on the parameter vector $\p$ (see~\cref{sec:linear_data_loss}). The construction follows standard tensor-product B-spline interpolation with prescribed knot vectors on $[0,1]$ (cf.~\cite{d_boor_practical_1978,butterfield_computation_1976}).

\begin{remark}[Equivalent spline representations]
The spline surface admits an equivalent representation in terms of the interpolation values,
\begin{align}
  \widehat{\W}(\xiu,\xiv)
  =
  \sum_{p=1}^{P} \theta_p\, S_p(\xiu,\xiv),
  \qquad
  S_p(\xiu,\xiv)=\frac{\partial \widehat{\W}}{\partial \theta_p},
\end{align}
where $S_p$ denotes the spline obtained by perturbing a single interpolation value. This representation is mathematically equivalent to the classical control-point formulation~\eqref{eq:surface}.

In the present work, we adopt the interpolation-based formulation, as it provides a direct physical interpretation of the parameters as energy values and enables a particularly transparent construction of linear constraints.
\end{remark}

\paragraph{Modeling assumptions and constraints}

The proposed spline representation satisfies the standard requirements of hyperelastic constitutive models by construction or through simple parameter constraints.

\begin{itemize}
  \item \emph{Objectivity:}  
  The strain-energy density is formulated in terms of the invariants $\Ione$ and $\Itwo$ of the isochoric right Cauchy--Green tensor. Consequently, $\W(\F)$ is frame-indifferent by construction.
  
  \item \emph{Stress-free reference configuration:}  
  For the chosen invariant formulation, the derivatives with respect to the deformation gradient vanish at the undeformed state,
  \begin{align}
    \frac{\partial \Ione}{\partial \F}\bigg|_{\F=\mathbf I} = \mathbf{0},
    \qquad
    \frac{\partial \Itwo}{\partial \F}\bigg|_{\F=\mathbf I} = \mathbf{0}.
  \end{align}
  Hence,
  \begin{align}
    \frac{\partial \W}{\partial \F}\bigg|_{\F=\mathbf I} = \mathbf{0},
  \end{align}
  and the nominal stress satisfies $\Pnom=\mathbf{0}$ at $\F=\mathbf I$ without requiring any constraints.
  
  \item \emph{Energy normalization:}  
  The condition $\W(\mathbf I)=0$ is enforced directly by fixing the interpolation value associated with the undeformed configuration. This reduces to a simple bound or equality constraint on a single parameter.
  
  \item \emph{Monotonicity and convexity:}  
  Physically motivated constraints on the strain-energy density are imposed through linear inequality constraints on the spline parameters. In the separable case, this yields monotone and convex (and thus polyconvex) models. For spline surfaces, monotonicity and directional convexity are enforced; see ~\cref{sec:constraints}.

\end{itemize}

With these properties established, the derivatives of the spline surface with respect to the mapped coordinates are
\begin{align}
  \widehat{\W}_{,\xiu}
  &=
  \sum_{i=1}^{n_\xiu}\sum_{j=1}^{n_\xiv}
  c_{ij}\,\Nbu{i}'(\xiu)\,\Nbv{j}(\xiv),
  \\
  \widehat{\W}_{,\xiv}
  &=
  \sum_{i=1}^{n_\xiu}\sum_{j=1}^{n_\xiv}
  c_{ij}\,\Nbu{i}(\xiu)\,\Nbv{j}'(\xiv).
\end{align}

Insertion into \cref{eq:dWd1,eq:dWd2} yields the invariant derivatives $W_{,1}$ and $\W_{,2}$ and therefore the analytical stresses \cref{eq:Put,eq:Pbt}.

\subsection{Constrained linear least-squares calibration problem}

Let the experimental dataset be
\begin{align}
  \calD=\calD_{\ut}\cup\calD_{\bt}\cup\calD_{\ps},
\end{align}
where each subset contains measured pairs $(\lambda_q,P_q^{\exp})$ corresponding to the deformation modes $\ut$, $\bt$, and $\ps$.

The unknown model parameters are the spline interpolation values collected in the vector
\begin{align}
  \p=\operatorname{vec}(\Ivec).
\end{align}

\paragraph{Data misfit}

The discrepancy between model predictions and experimental measurements
is measured by the weighted least-squares objective
\begin{align}
  \calL_{\mathrm{data}}(\p)
  &=
  \frac{1}{N_{\ut}}
  \sum_{q=1}^{N_{\ut}}
  \big[
  P_{\ut}^{\mathrm{mod}}(\lambda_q;\p)
  -
  P_{\ut,q}^{\exp}
  \big]^2
  \nonumber\\
  &\quad+
  \frac{1}{N_{\bt}}
  \sum_{q=1}^{N_{\bt}}
  \big[
  P_{\bt}^{\mathrm{mod}}(\lambda_q;\p)
  -
  P_{\bt,q}^{\exp}
  \big]^2
  \nonumber\\
  &\quad+
  \frac{1}{N_{\ps}}
  \sum_{q=1}^{N_{\ps}}
  \big[
  P_{\ps}^{\mathrm{mod}}(\gamma_q;\p)
  -
  P_{\ps,q}^{\exp}
  \big]^2 .
\end{align}

As shown in ~\cref{sec:linear_data_loss}, the predicted nominal stresses depend linearly on the interpolation values. Consequently, the residual vector can be written as
\begin{align}
  \bm r_{\mathrm{data}}(\p)
  =
  \mathbf{A}\p-\bm{y},
\end{align}

where $\bm{y}$ denotes the vector of experimental nominal stresses.

\paragraph{Regularization}

To stabilize spline parameters located in weakly sampled regions of invariant space, we add a curvature regularization term
\begin{align}
  \calR(\p)= \lambda_{\mathrm{pen}} \,\|\mathbf{A}_{\mathrm{pen}}\p\|_2^2 ,
\end{align}
where the penalty operator $\mathbf{A}_{\mathrm{pen}}$ is constructed from directional second derivatives of the spline surface, cf.~\cref{sec:curvature_penalty}. The regularization weight $\lambda_{\mathrm{pen}}$ controls the trade-off between data fidelity and suppression of spurious curvature in weakly sampled regions of invariant space. It is automatically selected by means of an L-curve analysis \cite{hansen_adaptive_2007,hansen_use_1993,cultrera_simple_2020}. Details of the procedure are given in ~\cref{sec:lcurve_lambda}.

\paragraph{Physical constraints}

Basic physical properties of the strain-energy function, such as monotonicity and directional convexity, are enforced through linear inequality constraints
\begin{align}
  \mathbf{A}_{\mathrm{ineq}}\p\le\mathbf{b}_{\mathrm{ineq}},
\end{align}
whose construction is described in ~\cref{sec:constraints}.

\paragraph{Optimization problem}

Combining the data term and the regularization yields the quadratic optimization problem
\begin{align}
  \min_{\p}
  \quad
  \|\mathbf{A}\p-\bm{y}\|_2^2
  +
  \lambda_{\mathrm{pen}} \|\mathbf{A}_{\mathrm{pen}}\p\|_2^2
  \quad
  \text{s.t.}
  \quad
  \mathbf{A}_{\mathrm{ineq}}\p\le\mathbf{b}_{\mathrm{ineq}}.
\end{align}

Because both the stress prediction and the regularization operator are linear in the interpolation values, the calibration reduces to a constrained linear least-squares problem. The framework is implemented in MATLAB. In particular, the spline construction utilizes the \texttt{spapi} function, while calibration is performed using the \texttt{lsqlin} optimization algorithm.

\section{Results and discussion}
\label{sec:results}

The present setting naturally suggests the following model classes:
\begin{align}
  \text{separable:}\quad 
  & \W(\Ione,\Itwo)=\W_1(\Ione)+\W_2(\Itwo), \\[4pt]
  \text{surface $(\Ione,\Itwo)$:}\quad 
  & \W(\Ione,\Itwo)=\Wsurf(\Ione,\Itwo), \\[4pt]
  \text{surface $(\xiu,\xiv)$:}\quad 
  & \W(\Ione,\Itwo)=\Wsurf\big(\xiu(\Ione),\xiv(\Ione,\Itwo)\big).
\end{align}

The first model retains both invariants but excludes coupling between them.  
The second introduces a tensor-product spline surface directly in the $(\Ione,\Itwo)$ space, thereby allocating interpolation support also to inadmissible regions.  
The third incorporates the proposed coordinate mapping and defines the spline surface on the unit square $(\xiu,\xiv)\in[0,1]^2$. Its pullback corresponds exactly to the admissible invariant domain, so that all spline support points are concentrated on physically realizable deformation states.

\medskip

To compare the different models, we consider the classical Treloar data set \cite{treloar_stress-strain_1944} \footnote{The Treloar experimental data used in this study were obtained from the supplementary material accompanying the work of ~\cite{dammas_when_2025}.}. This benchmark comprises the deformation modes $\ut$, $\bt$, and $\ps$ for representative hyperelastic material behavior. All deformation modes are jointly calibrated in the present study.

\medskip

When extending from univariate to bivariate spline representations, the coverage of the invariant space becomes critical. The considered deformation modes $\ut$, $\bt$, and $\ps$ generate only one-dimensional curves in the $(\Ione,\Itwo)$-plane, so that large regions remain unsampled.

To quantify parameter activation, we analyze the Jacobian matrix of the residual vector with respect to the model parameters. Let
\begin{align}
  \mathbf{r}(\p) =
  \begin{bmatrix}
  r_1(\p) \\
  \vdots \\
  r_N(\p)
  \end{bmatrix}
\end{align}
denote the vector of residuals between model predictions and experimental measurements, where $\p$ collects the spline parameters. The Jacobian matrix reads
\begin{align}
  \mathbf{J}(\p)
  =
  \frac{\partial \mathbf{r}}{\partial \p}
  \in \mathbb{R}^{N\times P}.
\end{align}

A convenient measure for parameter activation $j$ is the column norm
\begin{align}
  a_j
  =
  \left\| \mathbf{J}_{:,j} \right\|_2,
  \label{eq:activation}
\end{align}
which quantifies its influence on the residuals. Since each spline parameter corresponds to an interpolation point in invariant space, these values can be visualized as a heatmap over the spline grid.

For visualization purposes, we plot the relative activation
\begin{align}
  \tilde a_j
  =
  \frac{a_j}{\max_k a_k},
  \label{eq:rel_act}
\end{align}
using a logarithmic color scale $\log_{10}(\tilde a_j)$. Values close to zero indicate strongly activated parameters, whereas increasingly negative values correspond to weakly activated regions of the invariant space. Such regions typically require mild regularization of the spline surface, for example, through a curvature penalty in the objective function.

To assess the quality of fit, we report the mean squared error (MSE) and the coefficient of determination $R^2$.

\subsection{Separable model}

\begin{figure}
    \centering
    \includegraphics[width=0.9\linewidth]{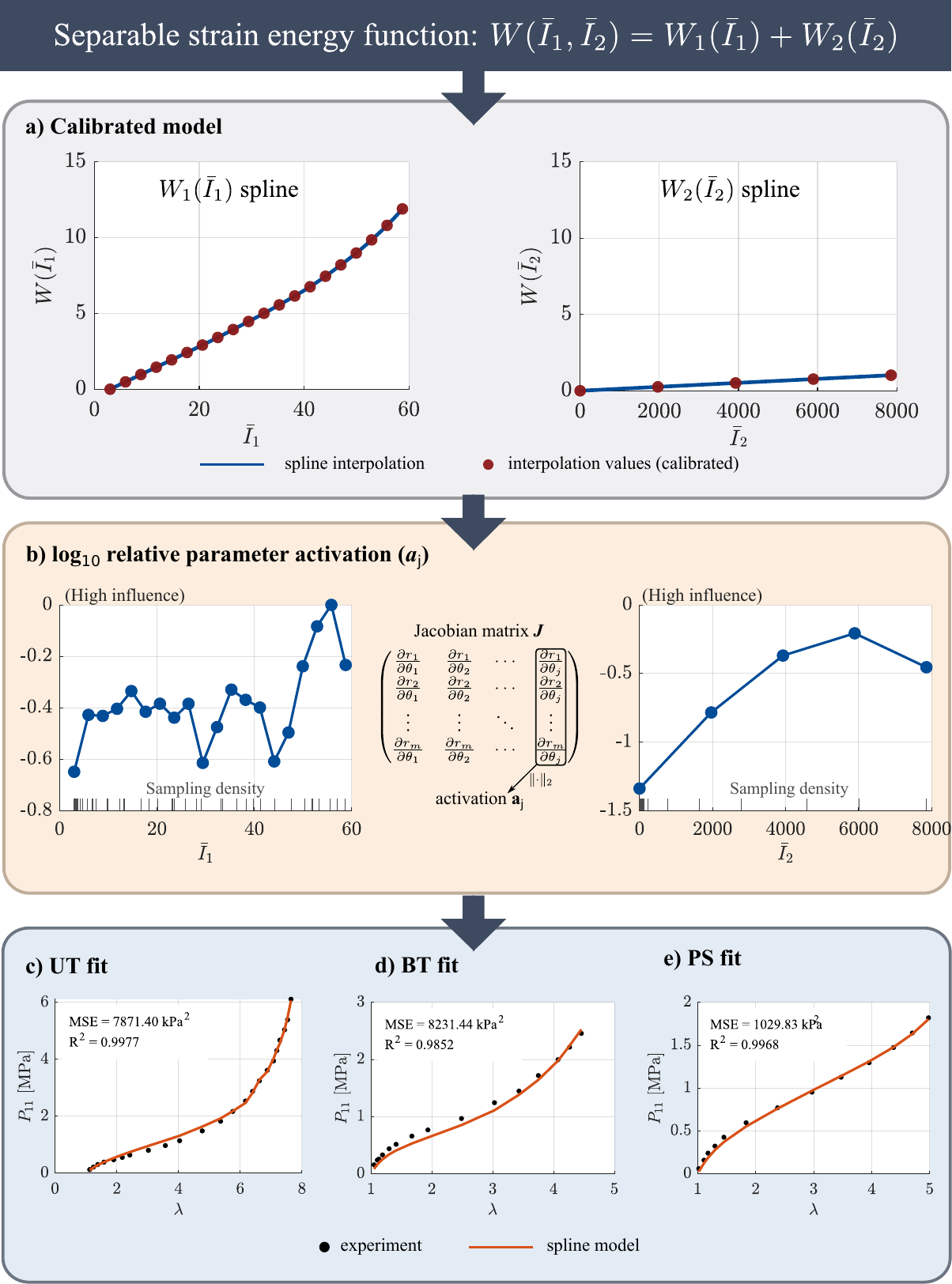}
    \caption{Results for the separable spline model. (a) Calibrated univariate splines $W_1(\Ione)$ and $W_2(\Itwo)$. (b) Logarithmic relative parameter activation as defined in ~\cref{eq:rel_act}. (c) Comparison of model predictions based on the splines in (a) with experimental Treloar data for the deformation modes $\ut$, $\bt$, and $\ps$.}
    \label{fig:univariate}
\end{figure}

As a first reference, we investigate the separable representation. The corresponding results are shown in ~\cref{fig:univariate}. The calibrated splines $W_1(\Ione)$ and $W_2(\Itwo)$, together with the relative parameter activation, are displayed in ~\cref{fig:univariate}a--b. Red circular markers indicate the optimized spline parameters, while solid lines represent the resulting interpolants. Owing to the imposed monotonicity and convexity constraints, the separable strain-energy density satisfies standard polyconvexity conditions.

The spline $W_2(\Itwo)$ remains nearly linear on its domain, whereas $W_1(\Ione)$ is initially linear and develops noticeable curvature towards the right end of the domain. The interpolation domains, $\Ione\approx\num{60}$ and $\Itwo\approx\num{8000}$, are chosen such that the largest invariant values from the calibration data lie close to the domain boundaries. The region where $W_1(\Ione)$ exhibits curvature coincides with the highest parameter activation, indicating that the spline adapts primarily in the regions of invariant space visited by the data.

Using the calibrated splines, we compare the predicted responses with the experimental measurements for all deformation modes; see \cref{fig:univariate}c. For $\ut$, $\bt$, and $\ps$, the mean squared errors are $\text{MSE}=\num{7871}\,\si{\kilo\pascal\squared}$, $\text{MSE}=\num{8231}\,\si{\kilo\pascal\squared}$, and $\text{MSE}=\num{1029}\,\si{\kilo\pascal\squared}$, respectively. 

Overall, the separable model captures the experimental response well despite the absence of invariant coupling. Nevertheless, allowing for coupled invariant dependence through a bivariate spline surface is expected to further improve the representation, which we investigate in the following section.

\subsection{Surface $(\Ione,\Itwo)$ model}

\begin{figure}
    \centering
    \includegraphics[width=0.9\linewidth]{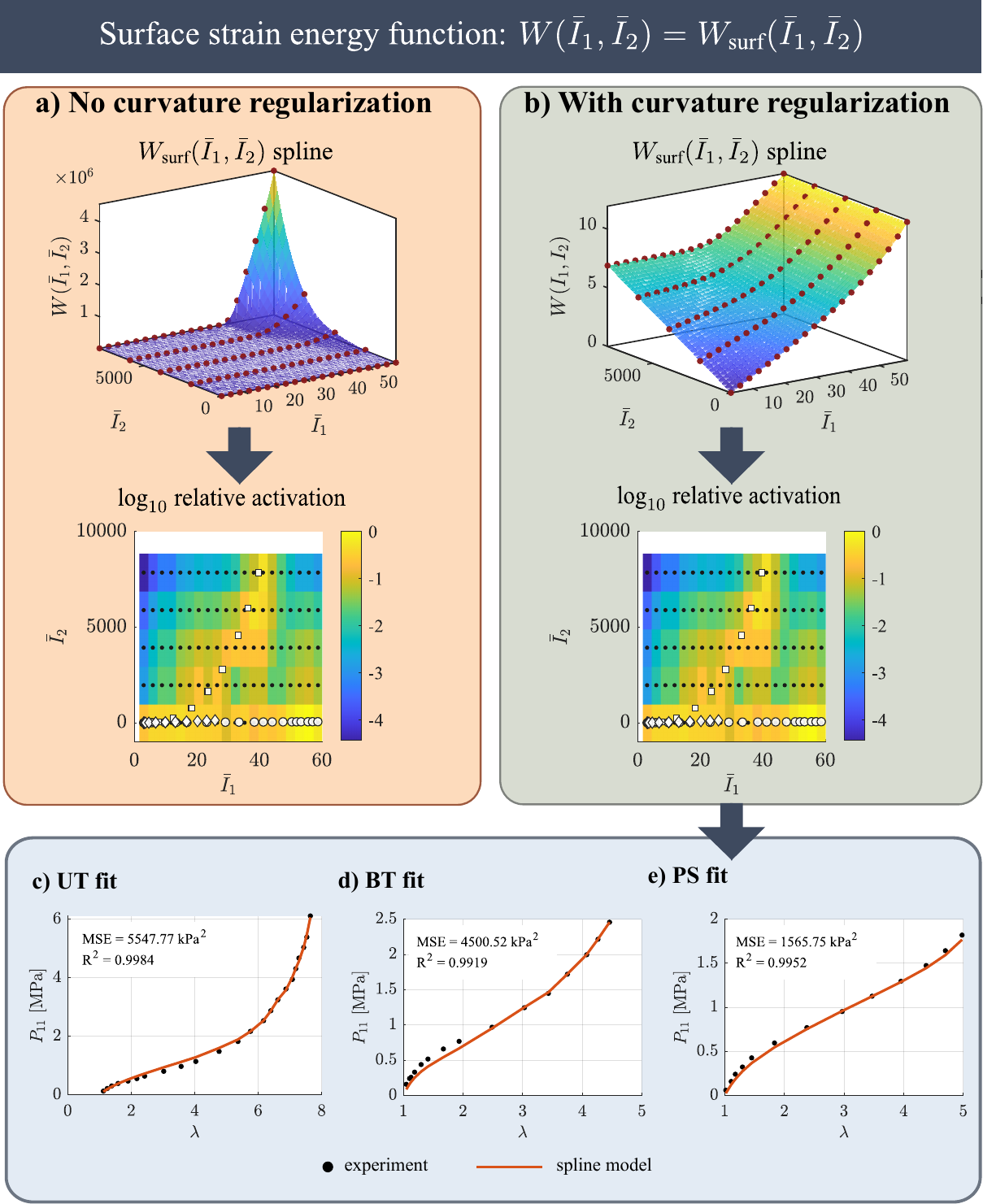}
    \caption{Results for the spline surface $\Wsurf(\Ione,\Itwo)$ in the $(\Ione,\Itwo)$ space. (a) Calibration without curvature regularization. (b) Calibration with a mild curvature penalty. The heatmap shows the logarithmic relative parameter activation as defined in ~\cref{eq:rel_act}. (c) Comparison of model predictions based on (b) with experimental Treloar data for the deformation modes $\ut$, $\bt$, and $\ps$.}
    \label{fig:surface_I1I2}
\end{figure}

The next extension of the separable model is a bivariate tensor-product spline surface defined directly in the $(\Ione,\Itwo)$ space. To enable a fair comparison with the separable model ($n_1 = 20$, $n_2 = 5$), we define $\num{20}$ and $\num{5}$ interpolation points in the respective invariant directions, resulting in $\num{100}$ model parameters.

We first calibrate the spline without curvature regularization. The resulting surface is shown in ~\cref{fig:surface_I1I2}a together with the heatmap of the log-scaled parameter activation. Red markers indicate the optimized interpolation values, while white markers in the heatmap represent the invariant tuples sampled by the deformation modes. The spline interpolation sites are overlaid for reference.

The first layer of the spline grid, corresponding to $\Itwo \approx 0$ \footnote{While $\Itwo \ge 3$ holds, the polyconvex invariant (cf.~\cref{eq:I2poly}) satisfies $\widetilde{\Itwo} \ge 0$.}, exhibits strong parameter activation driven by the $\ut$ mode, where large $\Ione$ values are attained. Another highly activated region appears along a vertical path near $\Ione \approx 40$, corresponding to the $\bt$ trajectory.

At the same time, several regions of the domain exhibit negligible activation. One example is the upper-left corner, characterized by small $\Ione$ and large $\Itwo$, which lies outside the physically admissible invariant domain and is therefore never sampled. A second region of weak activation occurs in the upper-right corner, where both invariants are large. Although this region is admissible, it is not visited by the Treloar deformation modes.

As a consequence, spline parameters in these regions remain essentially unconstrained. In the absence of regularization, they may attain extremely large values, reaching magnitudes on the order of $\num{e6}$. This reflects the ill-posedness of the calibration problem in unsampled regions rather than a physical effect.

To address this issue, we introduce a mild curvature penalty as defined in ~\cref{eq:constrained_problem}. The regularization weight $\lambda_{\mathrm{pen}}$ was selected via an L-curve analysis, yielding $\lambda_{\mathrm{pen}} = \num{3.6e-6}$ (cf. ~\cref{fig:curv_lambda} left). The resulting spline surface is shown in \cref{fig:surface_I1I2}b. While the parameter activation pattern remains unchanged, the unphysical spike in the unsampled region is eliminated, and the overall energy level becomes comparable to that of the separable model (cf.~\cref{fig:univariate}a).

Finally, we assess the gain in expressiveness due to coupled invariant dependence. The mean squared error decreases from $\num{7871}\,\si{\kilo\pascal\squared}$ to $\num{5547}\,\si{\kilo\pascal\squared}$ for $\ut$ and from $\num{8231}\,\si{\kilo\pascal\squared}$ to $\num{4500}\,\si{\kilo\pascal\squared}$ for $\bt$. For $\ps$, the error increases slightly from $\num{1029}\,\si{\kilo\pascal\squared}$ to $\num{1565}\,\si{\kilo\pascal\squared}$.

\subsection{Surface $(\xiu,\xiv)$ model}

\begin{figure}
    \centering
    \includegraphics[width=0.9\linewidth]{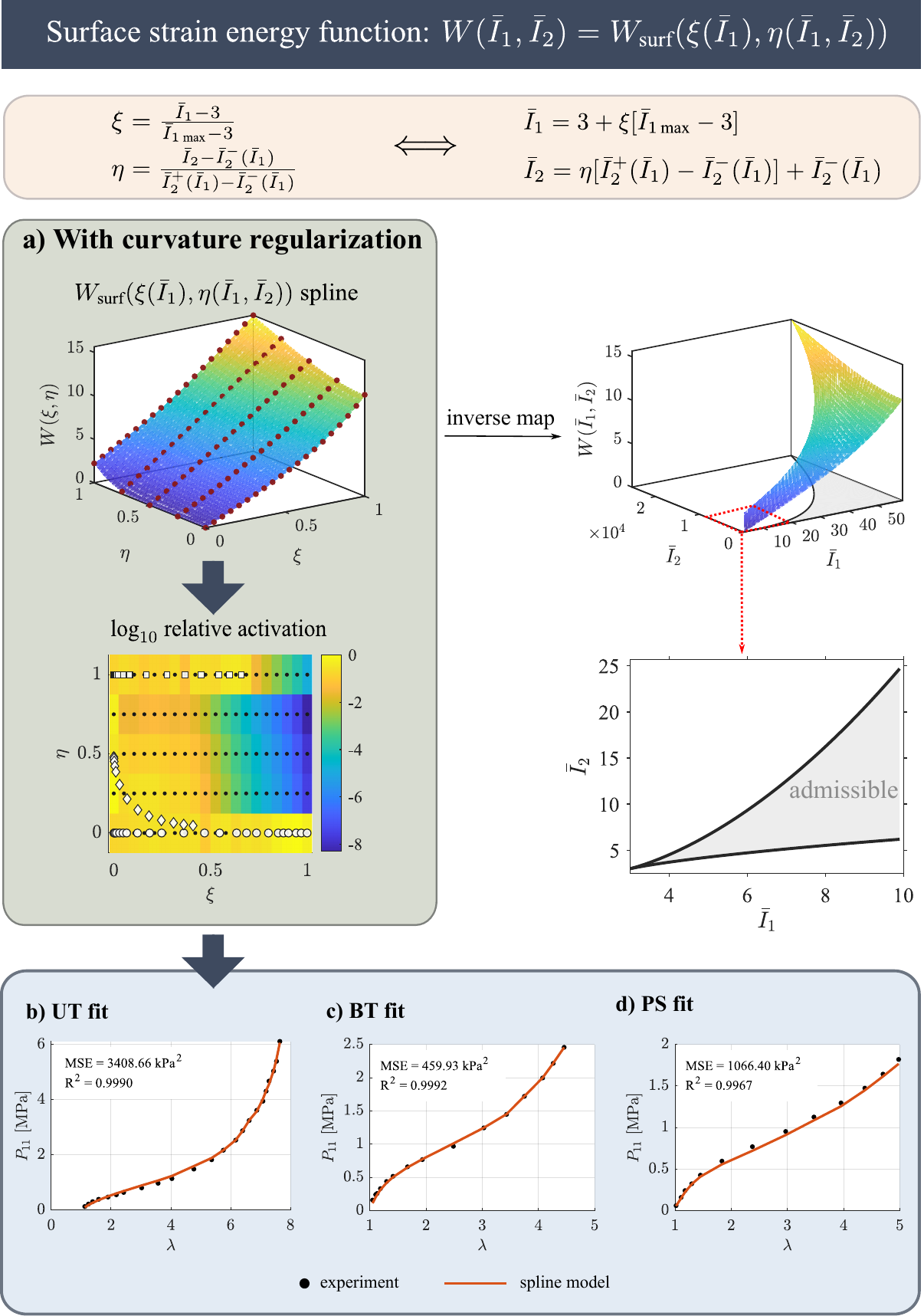}
    \caption{Results for the spline surface $\Wsurf(\xiu,\xiv)$ in normalized coordinates $(\xiu,\xiv)$. (a) Calibrated spline surface together with the logarithmic relative parameter activation. The inverse mapping of the spline grid from $(\xiu,\xiv)$ to $(\Ione,\Itwo)$ defines the same surface on the admissible invariant domain. (b) Comparison of model predictions based on (a) with experimental Treloar data for the deformation modes $\ut$, $\bt$, and $\ps$.}
    \label{fig:surface_adm}
\end{figure}

As demonstrated in the previous section, defining a bivariate spline surface directly in the $(\Ione,\Itwo)$ space is inefficient, as interpolation sites are allocated to regions outside the physically admissible invariant domain. To overcome this limitation, we consider the mapped model
\[
W(\xiu,\xiv),
\]
which utilizes the coordinate transformation introduced earlier and operates on the unit square interpolation domain $(\xiu,\xiv)\in[0,1]^2$.

The spline grid is discretized using $\num{20}$ interpolation points in the $\xiu$ direction and $\num{5}$ points in the $\xiv$ direction, resulting in $\num{100}$ parameters. Calibration is performed with a mild curvature penalty, motivated by the discussion in the previous section. The corresponding regularization weight $\lambda_{\mathrm{pen}} = \num{5.1e-6}$ was selected via an L-curve analysis (cf. ~\cref{fig:curv_lambda} right). In addition, a small regularization parameter $\num{1e-6}$ is introduced to avoid degeneracy of the mapping at the undeformed state, which we have found to have negligible influence on the model response.

The calibrated spline surface in $(\xiu,\xiv)$ space is shown in ~\cref{fig:surface_adm}a. Red markers denote the optimized spline interpolation values, while white markers indicate the invariant tuples sampled by the deformation modes. In this normalized coordinate system, these tuples admit a clear geometric interpretation. By construction, $\xiv=0$ corresponds to the lower boundary of the admissible invariant domain associated with uniaxial tension ($\ut$), and $\xiv=1$ corresponds to the upper boundary associated with equi-biaxial tension ($\bt$). The Treloar data, therefore, lie along the lines $\xiv=0$ and $\xiv=1$, while the pure shear data ($\ps$) occupy the lower-left portion of the grid.

These relations are directly reflected in the activation heatmap. Strong activation occurs along $\xiv=0$ and $\xiv=1$, as well as near $\xiu\approx0$, where the pure shear data are located. In contrast, weak activation is observed in the central region of the grid, roughly between $\xiv\approx0.25$ and $\xiv\approx0.75$, where no experimental data are available.

A complementary view is provided by the pull-back mapping to the physical $(\Ione,\Itwo)$ space, also shown in ~\cref{fig:surface_adm}a (right). The admissible invariant region is highlighted on the grid floor. The interpolation sites are not highlighted in this view to avoid suggesting that the optimization was performed directly in $(\Ione,\Itwo)$ space.

The key advantage of the $(\xiu,\xiv)$ formulation is that the same number of parameters is concentrated entirely on the physically admissible domain. In contrast to the $(\Ione,\Itwo)$ model, no degrees of freedom are wasted in inadmissible regions, leading to a more efficient use of the spline representation.

Finally, we assess the predictive performance of the model. The mean squared errors are $\num{3408},\si{\kilo\pascal\squared}$ for $\ut$, $\num{459},\si{\kilo\pascal\squared}$ for $\bt$, and $\num{1066},\si{\kilo\pascal\squared}$ for $\ps$. While the improvement is not uniform across all deformation modes, the mapped surface achieves a significantly lower combined error than both the separable and direct $(\Ione,\Itwo)$ surface models (cf. ~\cref{fig:performance_and_efficiency}a).

Finally, we note that, in contrast to the separable model, the spline surface formulations are constrained only through monotonicity and directional convexity and therefore do not guarantee strict polyconvexity. However, the curvature regularization effectively limits second derivatives, resulting in smooth and well-behaved energy landscapes. In all cases considered, the models exhibit stable and physically reasonable responses, indicating that the imposed constraints are sufficient for the present setting.

\subsection{A note on computational efficiency}

\begin{figure}
    \centering
    \includegraphics[width=0.98\linewidth]{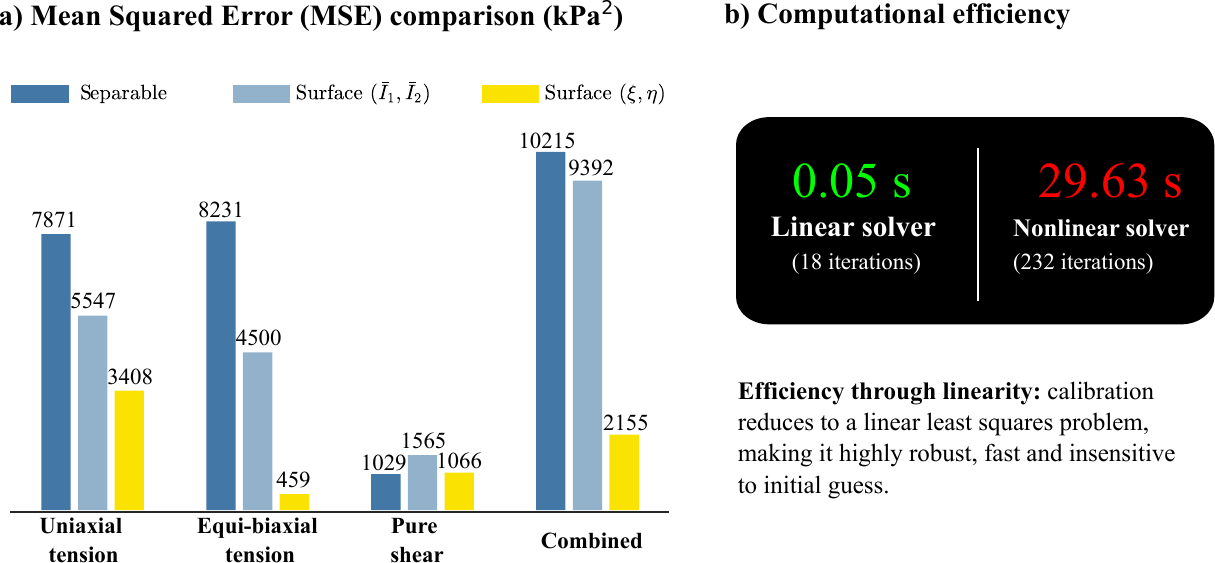}
    \caption{Quantitative evaluation of model accuracy and calibration time. (a) Mean squared error (MSE) for uniaxial tension, equi-biaxial tension, and pure shear. The combined error is defined as $\mathrm{MSE}_{\mathrm{combined}} = \big(\mathrm{MSE}_{\ut}^2 + \mathrm{MSE}_{\bt}^2 + \mathrm{MSE}_{\ps}^2\big)^{1/2}$. While the mapped $(\xiu,\xiv)$ surface does not achieve the lowest error for every individual mode (e.g., pure shear), it consistently yields the best overall performance in terms of the combined MSE.
(b) Calibration wall time. The linear dependence of the stress response on the spline parameters enables robust, near-instantaneous optimization that is insensitive to the initial guess.}
    \label{fig:performance_and_efficiency}
\end{figure}

As shown in ~\cref{sec:linear_data_loss}, the calibration of the spline interpolation values can be reformulated as a constrained linear least-squares problem. This holds whenever the kinematic quantities entering the stress relations (e.g., the deformation gradient) are fixed, as is the case for the homogeneous deformation states considered here.

To illustrate the computational implications, we compare two optimization strategies available in the MATLAB Optimization Toolbox: (i) the linear least-squares solver \texttt{lsqlin}, and (ii) the nonlinear least-squares solver \texttt{lsqnonlin}. All computations were performed on a workstation equipped with an Intel\textsuperscript{\textregistered} Core\texttrademark{} i7-13700 (13th Gen) processor.

As a benchmark, we consider the calibration of the previously introduced bivariate spline surface model with \num{100} model parameters. Using \texttt{lsqlin}, the optimization required \num{18} iterations with a wall time of \SI{0.05}{\second}. In contrast, \texttt{lsqnonlin} requires \num{232} iterations and \SI{29.63}{\second} to reach convergence (cf. ~\cref{fig:performance_and_efficiency}b).

In both cases, the resulting parameter vectors agree up to numerical precision. These results highlight that exploiting the linear structure of the problem yields orders-of-magnitude improvements in computational efficiency.

An additional advantage of the linear formulation is the robustness with respect to initializations. In particular, the interior-point variant of \texttt{lsqlin} does not require an initial guess, as the optimization problem is solved directly within the feasible region defined by the linear constraints.

\paragraph{Remark on solver efficiency}
Although the calibration problem is linear in the model parameters, a nonlinear least-squares solver such as \texttt{lsqnonlin} does not converge in a single iteration. This is because such solvers are designed for general nonlinear problems and therefore employ iterative strategies based on trust-region or damping techniques (e.g., Gauss--Newton or Levenberg--Marquardt methods). Even for linear residuals, these methods do not take the exact Newton step in a single iteration, but instead apply safeguarded updates whose acceptance is controlled by globalization strategies. Moreover, the presence of inequality constraints further prevents a one-step solution, as the active set must be identified iteratively.

\section{Conclusion}

We have presented a data-adaptive framework for constitutive identification in incompressible isotropic hyperelasticity based on a bivariate B-spline surface defined on the admissible invariant domain. By restricting attention to homogeneous deformation modes, calibration can be performed directly from analytical stress relations, eliminating the need for finite element model updating. Due to the linear dependence of the spline representation on its coefficients, the identification problem reduces to a constrained linear least-squares problem, enabling efficient and robust calibration.

The central idea is to align the approximation space with the physically admissible invariant domain. A direct tensor-product spline construction in the full $(\Ione,\Itwo)$ space allocates interpolation support to non-physical regions, leading to inefficient parameter usage and poorly constrained degrees of freedom. In contrast, the proposed mapping concentrates all spline parameters on physically realizable deformation states, resulting in a compact and effective representation.

The proposed formulation retains the interpretability of spline-based models while enabling explicit coupling between invariants. Compared to separable representations, it provides improved accuracy at comparable model complexity while preserving a robust and well-posed calibration procedure.

The linear structure and resulting computational efficiency make the framework particularly attractive for uncertainty quantification and Bayesian inference, where repeated model evaluations are required. In addition, the near-instantaneous calibration opens the door to real-time or interactive constitutive identification, for example, in conjunction with experimental testing.

\section*{Code availability}
The code for the calibration is publicly available in \url{https://github.com/swiesheier/bivariate-spline-hyperelasticity}.

\section*{Acknowledgments}
Simon Wiesheier, Miguel Angel Moreno-Mateos, and Paul Steinmann acknowledge support from the European Research Council (ERC) under the Horizon Europe program (Grant-No. 101052785, project: SoftFrac). Funded by the European Union. Views and opinions expressed are, however, those of the author(s) only and do not necessarily reflect those of the European Union or the European Research Council Executive Agency. Neither the European Union nor the granting authority can be held responsible for them. We also thank Franz Dammaß and Karl Kalina for the discussions in Erlangen that motivated this work.

\begin{figure}[ht]
\includegraphics[width=0.3\textwidth]{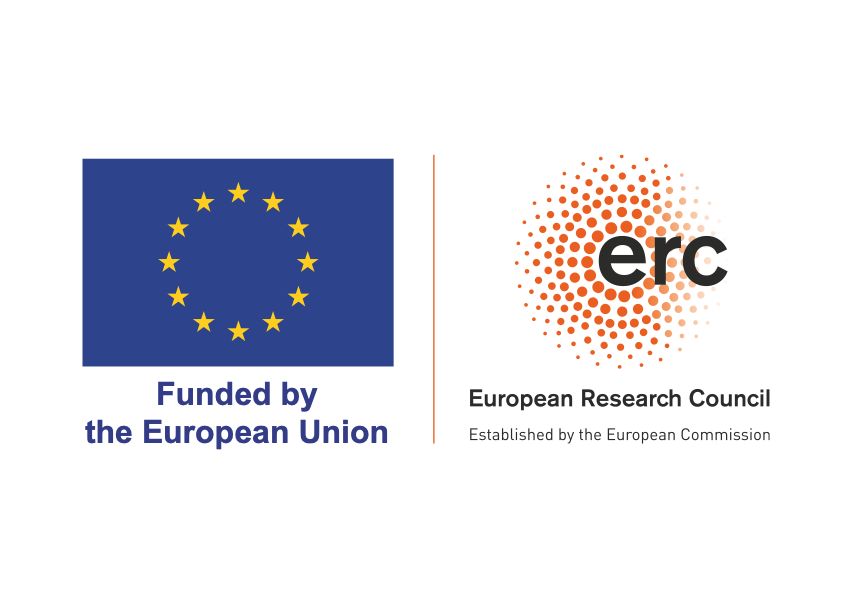}
\end{figure}

\section*{Competing Interests}
\noindent The Authors declare no Competing Financial or Non-Financial Interests.

\appendix

\section{Linearity of the data misfit}
\label{sec:linear_data_loss}

The spline surfaces used in the present work are constructed using interpolation splines. In MATLAB this is realized through the command \texttt{spapi}, which constructs a spline function from prescribed interpolation sites and interpolation values. The interpolation values constitute the model parameters to be identified.

Let the interpolation values be collected in the matrix
\begin{align}
  \Ivec \in \mathbb{R}^{n_\xiu \times n_\xiv},
\end{align}
and denote the parameter vector by
\begin{align}
  \p = \operatorname{vec}(\Ivec).
\end{align}

Given the interpolation sites and values, the spline construction
determines the B-spline control points through a linear system.
Consequently, the control points are linear functions of the
interpolation values. Since the spline surface is linear in the control
points, it follows that the strain-energy representation is also linear
in the interpolation values.

Therefore the strain energy can be written as
\begin{align}
  \W(\Ione,\Itwo)
  =
  \sum_{p=1}^{P} \theta_p\, S_p(\Ione,\Itwo),
\end{align}
where $P=n_\xiu n_\xiv$ and the functions
\begin{align}
  S_p(\Ione,\Itwo)
  =
  \frac{\partial W}{\partial \theta_p}
\end{align}
represent the parameter sensitivity splines associated with the
interpolation values.

Since differentiation is linear, the invariant derivatives entering the
stress relations satisfy
\begin{align}
  W_{,1}(\Ione,\Itwo)
  &=
  \sum_{p=1}^{P} \theta_p\,
  \frac{\partial S_p}{\partial \Ione},
  \\
  W_{,2}(\Ione,\Itwo)
  &=
  \sum_{p=1}^{P} \theta_p\,
  \frac{\partial S_p}{\partial \Itwo}.
\end{align}

For the homogeneous deformation modes considered in this work, the
nominal stress can be written in the generic form
\begin{align}
  P_{11}
  =
  \alpha(F)\,W_{,1}
  +
  \beta(F)\,W_{,2},
\end{align}
where $\alpha(F)$ and $\beta(F)$ depend only on the deformation
gradient $F$ associated with the loading protocol.

Substituting the linear expansions of $W_{,1}$ and $W_{,2}$ yields
\begin{align}
  P_{11}(F;\p)
  =
  \sum_{p=1}^{P}
  \theta_p\,
  \Big[
    \alpha(F)\frac{\partial S_p}{\partial \Ione}
    +
    \beta(F)\frac{\partial S_p}{\partial \Itwo}
  \Big].
\end{align}

Hence the predicted stress depends linearly on the interpolation
parameters,
\begin{align}
  P_{11}(F;\p)
  =
  \bm a(F)^\top \p ,
\end{align}
with coefficient vector $\bm a(F)$ determined by the deformation state.

Let $\{F_k,P_k^{\exp}\}_{k=1}^{N}$ denote the experimental deformation
states and corresponding measured stresses.
The residual associated with the $k$-th data point reads
\begin{align}
  r_k(\p)
  =
  P_{11}(F_k;\p) - P_k^{\exp}
  =
  \bm a(F_k)^\top \p - P_k^{\exp}.
\end{align}

Stacking all residuals yields the linear system
\begin{align}
  \bm r(\p)
  =
  \mathbf A \p - \bm y,
\end{align}
where
\begin{align}
  A_{k,:} = \bm a(F_k)^\top,
  \qquad
  y_k = P_k^{\exp}.
\end{align}

Consequently, the data misfit term of the calibration problem becomes
\begin{align}
  \|\bm r(\p)\|_2^2
  =
  \|\mathbf A \p - \bm y\|_2^2,
\end{align}
which is a linear least-squares problem in the interpolation values.

\section{Construction of the curvature penalty operator}
\label{sec:curvature_penalty}

To stabilize spline parameters located in weakly sampled regions of invariant space, we augment the calibration problem by a mild curvature regularization. For the spline surface
\begin{align}
  \W(\xiu,\xiv),
\end{align}
defined on the tensor-product interpolation grid, we consider the functional
\begin{align}
  \calR[\W]
  =
  \lambda_{\mathrm{pen}}
  \int_{\Omega}
  \left[
      \W_{\xiu\xiu}(\xiu,\xiv)^2 + W_{\xiv\xiv}(\xiu,\xiv)^2
  \right]
  \,\mathrm{d}\Omega,
  \label{eq:curvature_penalty}
\end{align}
where $\Omega$ denotes the spline domain and $\lambda_{\mathrm{pen}}>0$
controls the strength of the penalty.

The spline surface is parameterized by its interpolation values. Let
\begin{align}
  \p
  =
  \begin{bmatrix}
  \theta_1 & \dots & \theta_P
  \end{bmatrix}^{\top}
\end{align}
collect these values. As shown in ~\cref{sec:linear_data_loss}, the spline surface is linear in the interpolation values and can be written as
\begin{align}
  W(\xiu,\xiv)
  =
  \sum_{p=1}^{P}
  \theta_p\, S_p(\xiu,\xiv),
  \label{eq:surface_sensitivity_expansion}
\end{align}
where $S_p(\xiu,\xiv)$ denotes the parameter sensitivity surface associated
with the $p$-th interpolation value,
\begin{align}
  S_p(\xiu,\xiv)=\frac{\partial W(\xiu,\xiv)}{\partial \theta_p}.
\end{align}

\begin{figure}
    \centering
    \includegraphics[width=0.8\linewidth]{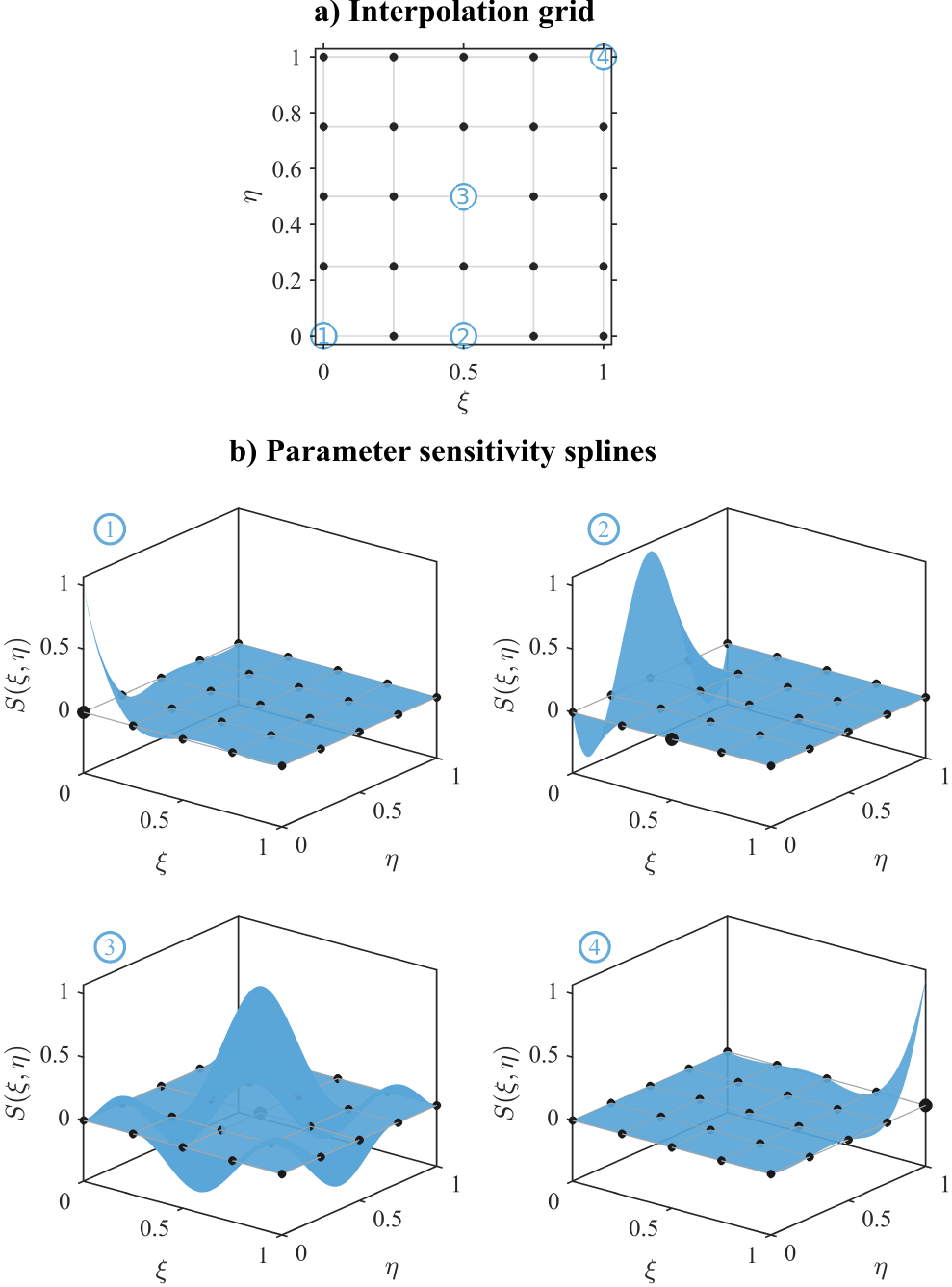}
    \caption{Parameter sensitivity surfaces of the tensor-product spline representation. (a) Interpolation grid in the mapped invariant coordinates. (b) Representative sensitivity functions associated with selected interpolation sites. These functions correspond to the columns of the linear operators used to enforce curvature regularization and convexity constraints.}
    \label{fig:sensitivty}
\end{figure}

Because differentiation is linear, the second derivatives of the spline
surface admit the representations
\begin{align}
  W_{\xiu\xiu}(\xiu,\xiv)
  &=
  \sum_{p=1}^{P}
  \theta_p\, S_{p,\xiu\xiu}(\xiu,\xiv),
  \\
  W_{\xiv\xiv}(\xiu,\xiv)
  &=
  \sum_{p=1}^{P}
  \theta_p\, S_{p,\xiv\xiv}(\xiu,\xiv).
\end{align}

Substituting these expressions into \eqref{eq:curvature_penalty}
yields a quadratic form in the interpolation values.

For numerical evaluation the spline domain is partitioned into knot spans and tensor-product Gauss quadrature is used. Let $(\xiu_q,\xiv_q)$ denote the quadrature points with weights $w_q$. Then
\begin{align}
  \calR[W]
  \approx
  \lambda_{\mathrm{pen}}
  \sum_q
  w_q
  \left[
      W_{\xiu\xiu}(\xiu_q,\xiv_q)^2
      +
      W_{\xiv\xiv}(\xiu_q,\xiv_q)^2
  \right].
\end{align}

Introducing the vectors
\begin{align}
  \mathbf{s}_{\xiu\xiu}(\xiu_q,\xiv_q)
  &=
  \begin{bmatrix}
  S_{1,\xiu\xiu}(\xiu_q,\xiv_q) & \dots & S_{P,\xiu\xiu}(\xiu_q,\xiv_q)
  \end{bmatrix}^{\top},
  \\
  \mathbf{s}_{\xiv\xiv}(\xiu_q,\xiv_q)
  &=
  \begin{bmatrix}
  S_{1,\xiv\xiv}(\xiu_q,\xiv_q) & \dots & S_{P,\xiv\xiv}(\xiu_q,\xiv_q)
  \end{bmatrix}^{\top},
\end{align}
we obtain
\begin{align}
  W_{\xiu\xiu}(\xiu_q,\xiv_q)
  &= \mathbf{s}_{\xiu\xiu}(\xiu_q,\xiv_q)^{\top}\p,
  \\
  W_{\xiv\xiv}(\xiu_q,\xiv_q)
  &= \mathbf{s}_{\xiv\xiv}(\xiu_q,\xiv_q)^{\top}\p .
\end{align}

Collecting all contributions yields the penalty operator
\begin{align}
  \mathbf{A}_{\mathrm{pen}}
  =
  \begin{bmatrix}
  \sqrt{w_1}\,\mathbf{s}_{\xiu\xiu}(\xiu_1,\xiv_1)^{\top} \\
  \sqrt{w_1}\,\mathbf{s}_{\xiv\xiv}(\xiu_1,\xiv_1)^{\top} \\
  \vdots \\
  \sqrt{w_{N_q}}\,\mathbf{s}_{\xiu\xiu}(\xiu_{N_q},\xiv_{N_q})^{\top} \\
  \sqrt{w_{N_q}}\,\mathbf{s}_{\xiv\xiv}(\xiu_{N_q},\xiv_{N_q})^{\top}
  \end{bmatrix},
\end{align}
so that
\begin{align}
  \mathcal{R}[W] \approx \lambda_{\mathrm{pen}} \|\mathbf{A}_{\mathrm{pen}}\p\|_2^2 .
\end{align}

\paragraph{Remark}
Regularization penalizes only the second directional derivatives  $W_{\xiu\xiu}$ and $W_{\xiv\xiv}$. This was found sufficient to suppress unphysical curvature in weakly sampled regions while preserving the flexibility of the spline surface in the data-relevant parts of the admissible invariant domain.

\section{Selection of the curvature regularization parameter}
\label{sec:lcurve_lambda}

The regularization parameter $\lambda_{\mathrm{pen}}$ controls the balance between data fidelity and curvature penalization. Its selection is critical, as too small values lead to spurious curvature of the spline surfaces in weakly sampled regions of the invariant space. In contrast, too large values degrade agreement with experimental data.

We determine $\lambda_{\mathrm{pen}}$ automatically using an L-curve analysis, a standard approach in the regularization of inverse problems \cite{hansen_adaptive_2007,hansen_use_1993,cultrera_simple_2020}. For a logarithmically spaced set of candidate values $\{\lambda_i\}_{i=1}^N$, we solve the constrained least-squares problem
\begin{align}
\min_{\p}\quad 
\|\mathbf{A}\p-\bm{y}\|_2^2 + \lambda_i\,\|\mathbf{A}_{\mathrm{pen}}\p\|_2^2
\quad \text{s.t.} \quad
\mathbf{A}_{\mathrm{ineq}}\p \le \mathbf{b}_{\mathrm{ineq}}.
\end{align}

For each solution $\p(\lambda_i)$, we record the pair
\begin{align}
\Big(\|\mathbf{A}\p(\lambda_i)-\bm{y}\|_2^2,\;
\|\mathbf{A}_{\mathrm{pen}}\p(\lambda_i)\|_2^2\Big),
\end{align}

and plot these quantities against each other in double-logarithmic scale. 

As $\lambda_{\mathrm{pen}}$ increases, the relative weight of the regularization term grows, leading to smoother spline surfaces and thus decreasing values of $\|\mathbf{A}_{\mathrm{pen}}\p(\lambda_i)\|_2^2$. At the same time, the agreement with the experimental data deteriorates, resulting in an increase of $\|\mathbf{A}\p(\lambda_i)-\bm{y}\|_2^2$. Conversely, for small values of $\lambda_{\mathrm{pen}}$, the optimization is dominated by the data misfit term, yielding highly accurate fits but potentially spurious curvature in weakly sampled regions. The resulting L-curve therefore exhibits a characteristic transition between a data-dominated regime and a regularization-dominated regime, with the corner indicating the balance between these competing effects.

Since the L-curve is available only at discrete values of $\lambda_i$, its corner is identified using a local geometric curvature measure in logarithmic coordinates. Defining
\begin{align}
x_i = \log \|\mathbf{A}\p(\lambda_i)-\bm{y}\|_2^2,
\qquad
y_i = \log \|\mathbf{A}_{\mathrm{pen}}\p(\lambda_i)\|_2^2,
\end{align}
we compute, for each interior point $i=2,\dots,N-1$, a discrete curvature indicator based on three consecutive points:
\begin{align}
\kappa_i = \frac{2\,\mathrm{area}\big((x_{i-1},y_{i-1}),(x_i,y_i),(x_{i+1},y_{i+1})\big)}
{\| \bm{p}_i - \bm{p}_{i-1} \| \, \| \bm{p}_{i+1} - \bm{p}_i \| \, \| \bm{p}_{i+1} - \bm{p}_{i-1} \|},
\end{align}
where $\bm{p}_i = (x_i,y_i)$. The corner index is then given by
\begin{align}
i^\ast = \arg\max_i \kappa_i,
\qquad
\lambda_{\mathrm{corner}} = \lambda_{i^\ast}.
\end{align}
This procedure provides a reproducible estimate of the point of maximum curvature of the Pareto front.

While the discrete curvature criterion provides a robust estimate of the corner of the L-curve, it is consistently observed that the corresponding parameter value $\lambda_{\mathrm{corner}}$ lies slightly within the regime where the data misfit begins to increase more rapidly. We therefore select a value of $\lambda_{\mathrm{pen}}$ slightly smaller than $\lambda_{\mathrm{corner}}$. In particular, we define
\begin{align}
\lambda_{\mathrm{pen}} = 10^{-1}\,\lambda_{\mathrm{corner}}.
\end{align}
This choice ensures that the solution remains in the regime where the data misfit is still close to its minimum, while the regularization term already provides sufficient smoothing of the spline surface.

\begin{figure}
    \centering
    \includegraphics[width=0.8\linewidth]{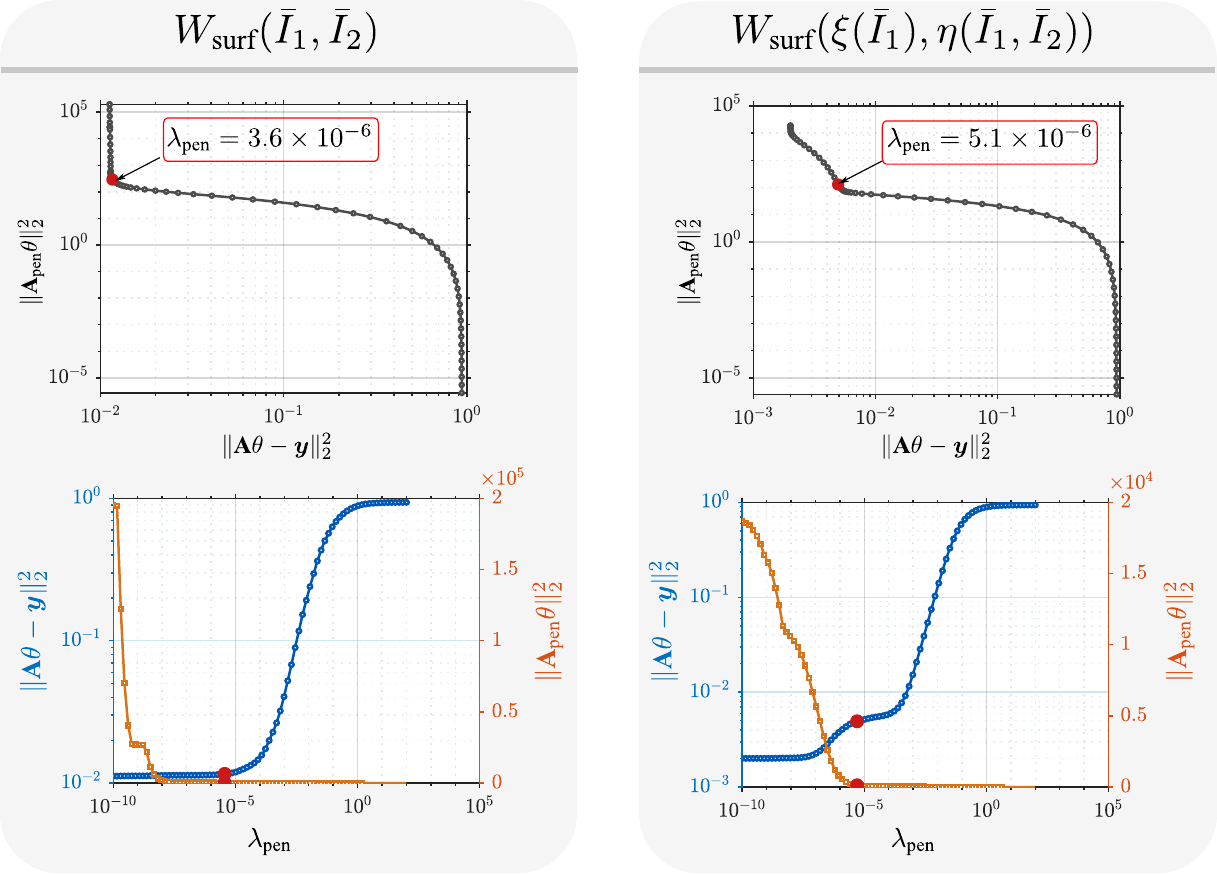}
    \caption{Automatic selection of the curvature regularization parameter $\lambda_{\mathrm{pen}}$ by means of an L-curve analysis for the $\Wsurf(\Ione,\Itwo)$ (left) and $\Wsurf(\xiu(\Ione),\xiv(\Ione,\Itwo))$ (right) model.}
    \label{fig:curv_lambda}
\end{figure}

The Pareto curves for the $\Wsurf(\Ione,\Itwo)$ and $\Wsurf(\xiu(\Ione),\xiv(\Ione,\Itwo))$ model are shown in ~\cref{fig:curv_lambda}.

\section{Construction of linear inequality constraints}
\label{sec:constraints}

The calibration problem introduced in the main text is formulated as
\begin{align}
\min_{\p}
\quad
\|\mathbf{A}\p-\bm{y}\|_2^2
+
\lambda_{\mathrm{pen}}\|\mathbf{A}_{\mathrm{pen}}\p\|_2^2
\qquad
\text{subject to}
\qquad
\mathbf{A}_{\mathrm{ineq}}\p
\le
\mathbf{b}_{\mathrm{ineq}} ,
\label{eq:constrained_problem}
\end{align}
where $\p$ collects the spline interpolation values. The matrix
$\mathbf{A}_{\mathrm{ineq}}$ is used to enforce monotonicity and convexity
properties of the spline representation.

\subsection*{Sensitivity spline representation}

As shown in ~\cref{sec:linear_data_loss}, spline interpolation is
linear in the interpolation values. Hence a generic spline in the variable $x$ can be written as
\begin{align}
W(x) = \sum_{p=1}^{P} \theta_p\, S_p(x),
\end{align}
where $S_p(x)$ denotes the spline obtained by perturbing the $p$-th
interpolation value by one unit. These functions will be referred to as
\emph{parameter sensitivity splines}. By definition,
\begin{align}
S_p(x)=\frac{\partial W(x)}{\partial \theta_p}.
\end{align}

Since differentiation is linear, derivatives of the spline again admit a linear
representation in terms of the parameters,
\begin{align}
W'(x) &= \sum_{p=1}^{P} \theta_p\, S_p'(x),\\
W''(x) &= \sum_{p=1}^{P} \theta_p\, S_p''(x).
\end{align}

Consequently, monotonicity and convexity requirements can be translated into
linear inequality constraints in the parameter vector $\p$.

\subsection*{Univariate case}

Consider a univariate spline $W(x)$ defined on interpolation sites
\begin{align}
x=\{x_1,\dots,x_P\}.
\end{align}
The parameter sensitivity splines $S_p(x)$ are defined as the spline functions
obtained by interpolating the canonical unit vectors on the same grid, i.e.,
\begin{align}
S_p(x_i)=\delta_{ip}, \qquad i=1,\dots,P,
\end{align}
where $\delta_{ip}$ denotes the Kronecker delta.

Each sensitivity spline admits a coefficient representation
\begin{align}
S_p(x)=\sum_{j} c_{pj} N_j(x),
\end{align}
where $N_j(x)$ are the B-spline basis functions. Collecting the coefficients
(control points) of all sensitivity splines yields a matrix
\begin{align}
\mathbf{C}=
\begin{bmatrix}
c_{11} & \dots & c_{1P}\\
\vdots &       & \vdots\\
c_{P1} & \dots & c_{PP}
\end{bmatrix}.
\end{align}

Using this representation, the derivative of the spline can be written as
\begin{align}
W'(x)
=
\sum_{p=1}^{P} \theta_p S_p'(x)
=
\sum_{j}
\Big[
\sum_{p=1}^{P} \theta_p\, c_{pj}^{(1)}
\Big]
N_j(x),
\end{align}
where $c_{pj}^{(1)}$ denote the coefficients of $S_p'(x)$ in the B-spline basis.

Since B-spline basis functions are non-negative and form a partition of unity,
a sufficient condition for
\begin{align}
W'(x)\ge 0 \quad \forall x
\end{align}
is that all corresponding coefficients are non-negative, i.e.,
\begin{align}
\sum_{p=1}^{P} \theta_p\, c_{pj}^{(1)} \ge 0 \quad \forall j.
\end{align}
This condition can be written in matrix form as
\begin{align}
-\mathbf{C}^{(1)}\p \le \mathbf 0 .
\end{align}

An analogous argument applies to the second derivative. Convexity,
\begin{align}
W''(x)\ge 0,
\end{align}
is ensured by requiring non-negativity of the corresponding coefficients,
leading to
\begin{align}
-\mathbf{C}^{(2)}\p \le \mathbf 0 .
\end{align}

Stacking these relations produces the matrix
\begin{align}
\mathbf{A}_{\mathrm{ineq}}=
\begin{bmatrix}
-\mathbf{C}^{(1)}\\
-\mathbf{C}^{(2)}
\end{bmatrix},
\qquad
\mathbf{b}_{\mathrm{ineq}}=\mathbf{0}.
\end{align}

Related constructions based on constrained spline parameterizations have recently been considered in neural-network architectures, for example, in the context of input-convex Kolmogorov--Arnold networks \cite{thakolkaran_can_2025}. In these approaches, spline functions are parameterized directly through their control points, and monotonicity or convexity is enforced via inequalities on the control points and their finite differences.

In the present work, by contrast, the unknowns are the interpolation values defining the spline through an interpolation operator. As a consequence, the constraint construction is formulated in terms of parameter sensitivity splines, leading to linear inequalities directly in the interpolation-value vector. While both approaches rely on the same underlying properties of B-spline bases, the resulting constraint operators differ due to the chosen parameterization. The use of interpolation values is particularly convenient in the present setting, as it provides a direct and interpretable link between the parameters and the values of the strain-energy function at physically meaningful points in invariant space.

\subsection*{Spline surface case}

For spline surfaces, the same construction applies. The interpolation values are
arranged on a tensor-product grid
\begin{align}
\{(\xiu_i,\xiv_j)\}_{i=1,\dots,n_\xiu;\,j=1,\dots,n_\xiv},
\end{align}
with $P=n_\xiu n_\xiv$ interpolation values.

Parameter sensitivity surfaces are obtained by interpolating canonical unit
perturbations on this grid. In particular, the $p$-th sensitivity surface
corresponds to interpolation data in which exactly one grid point is assigned
the value one and all others are zero.

Accordingly, the spline surface admits the representation
\begin{align}
W(\xiu,\xiv)=\sum_{p=1}^{P} \theta_p\, S_p(\xiu,\xiv),
\end{align}
where $S_p(\xiu,\xiv)$ denote the parameter sensitivity surfaces, i.e.,
\begin{align}
S_p(\xiu,\xiv)=\frac{\partial W(\xiu,\xiv)}{\partial \theta_p}.
\end{align}

Directional derivatives again admit a linear representation,
\begin{align}
W_\xiu(\xiu,\xiv) &= \sum_{p=1}^{P} \theta_p\, S_{p,\xiu}(\xiu,\xiv),\\
W_\xiv(\xiu,\xiv) &= \sum_{p=1}^{P} \theta_p\, S_{p,\xiv}(\xiu,\xiv),
\end{align}
and analogously for second derivatives.

Each derivative can be expressed in the tensor-product B-spline basis,
\begin{align}
W_\xiu(\xiu,\xiv)
=
\sum_{i,j} \alpha_{ij}\, N_i(\xiu) M_j(\xiv),
\end{align}
with coefficients $\alpha_{ij}$ that depend linearly on the parameter vector $\p$.

Since the tensor-product B-spline basis functions are non-negative and form a
partition of unity, non-negativity of the derivatives is ensured by requiring
non-negativity of the corresponding coefficients. This leads to linear inequality
constraints of the form
\begin{align}
\mathbf{A}_{\mathrm{ineq}}\p \le \mathbf 0 .
\end{align}

In the present work, we enforce directional monotonicity and convexity via
\begin{align}
W_\xiu(\xiu,\xiv)\ge0,\qquad
W_\xiv(\xiu,\xiv)\ge0,\qquad
W_{\xiu\xiu}(\xiu,\xiv)\ge0,\qquad
W_{\xiv\xiv}(\xiu,\xiv)\ge0.
\end{align}

%

%

\paragraph{Remark}
The surface model enforces convexity only in the coordinate directions.
Full two-dimensional convexity would require the Hessian of $W$ to be
positive semidefinite, which introduces nonlinear constraints and is therefore
not considered here.

\bibliographystyle{naturemag}

\end{document}